\documentstyle[prd,aps,twocolumn]{revtex}
\begin{document}
\draft

\def \d {{\rm d}}

\title{Uniformly accelerating black holes in a de Sitter universe}

\author{J. Podolsk\'y
 \thanks{E--mail: {\tt Podolsky@mbox.troja.mff.cuni.cz}}}
\address{Institute of Theoretical Physics, Charles University,
V Hole\v{s}ovi\v{c}k\'ach 2, 18000 Prague 8, Czech Republic.}

\author{J. B. Griffiths
 \thanks{E--mail: {\tt J.B.Griffiths@Lboro.ac.uk}}}
\address{Department of Mathematical Sciences,
Loughborough University,
Loughborough, Leicestershire LE11 3TU, U.K.}

\date{\today}

\maketitle
\begin{abstract}
A class of exact solutions of Einstein's equations is analysed which
describes uniformly accelerating charged black holes in an asymptotically
de~Sitter universe. This is a generalisation of the C-metric which
includes a cosmological constant. The physical interpretation of the
solutions is facilitated by the introduction of a new coordinate system
for de~Sitter space which is adapted to accelerating observers in this
background. The solutions considered reduce to this form of the de~Sitter
metric when the mass and charge of the black holes vanish. 
\end{abstract}


\narrowtext

\section{Introduction}

It has recently been suggested \cite{ManRos95} that the pair creation of
black holes is possible in a background with a cosmological constant
as this supplies the necessary negative potential energy. To investigate
this process, an exact solution of Einstein's equations is required which
represents a pair of black holes which accelerate away from each other in
a de~Sitter background. In fact, a general class of such solutions is
available \cite{PleDem76} (and used in \cite{ManRos95})  which includes
mass and charge parameters as well as a cosmological constant $\Lambda$.
When \ $\Lambda=0$, \ these solutions include the well-known C-metric
which describes two black holes which uniformly accelerate in a Minkowski
background under the action of conical singularities \cite{KinWal70}. 
However, for \ $\Lambda\ne0$, \ the physical, geometrical and global
properties of the space-time \cite{PleDem76} has not been investigated
thoroughly even at the classical level.

It is the purpose of the present paper to provide a physical
interpretation for this family of exact solutions. We will show that
indeed it represents (possibly charged) black holes uniformly accelerating
in a de~Sitter universe. This interpretation is obtained after first
introducing a new coordinate system which is adapted to the motion of
(two) uniformly accelerating test particles in de~Sitter space. We will
then show that, for small mass and charge parameters, some of the
Plebanski--Demianski solutions \cite{PleDem76} can be regarded as
perturbations of de~Sitter space in these coordinates. This enables the
physical meaning of the parameters to be determined.

\section{Uniformly accelerating observers in de Sitter space}

It is well-known that de~Sitter space can be represented as a 4-hyperboloid
  \begin{eqnarray}
 -Z_0^2+Z_1^2+Z_2^2+Z_3^2+Z_4^2=a^2\equiv3/\Lambda
 \label{fourhyp}
  \end{eqnarray}
 in a 5-dimensional
Minkowski space.  In the familiar parametrization by static coordinates,
  \begin{eqnarray}
 &&Z_0=\sqrt{a^2-R^2}\,\sinh(T/a) \>, \nonumber\\
 &&Z_1=\pm\sqrt{a^2-R^2}\,\cosh(T/a) \>, \nonumber\\
 &&Z_4=R\cos\Theta \>, \label{static}\\
 &&Z_2=R\sin\Theta\cos\Phi \>, \nonumber\\
 &&Z_3=R\sin\Theta\sin\Phi \>, \nonumber
 \end{eqnarray}
 the metric is,
 \begin{equation}
 \d s^2 ={\d R^2\over1-{R^2\over a^2}} +R^2(\d
\Theta^2+\sin^2\Theta\,\d\Phi^2) -{\textstyle \left(1-{R^2\over
a^2}\right)\d T^2} ,
\label{staticmetr}
 \end{equation}
 with \ $R\in[0,a]$, \ $T\in(-\infty,\infty)$, \
$\Theta\in[0,\pi]$, \ $\Phi\in[0,2\pi)$. \
The two maps given by (\ref{static}) cover two causally disconnected areas
\ $Z_1>0$ \ and \ $Z_1<0$ \ of the manifold.

Now, let us consider the timelike worldlines $x^\mu(\tau)$:
\begin{equation}
R=R_0\>,\  \Theta=\Theta_0\>,\
\Phi=\Phi_0\>, \  T=a\tau/\sqrt{a^2-R_0^2}\>,
\label{worldlines}
\end{equation}
where $R_0, \Theta_0$ and $\Phi_0$ are constants. The 4-velocity is \ 
$u^\mu=(0,0,0,a/\sqrt{a^2-R_0^2})$, \ and the 4-acceleration is
$\dot u^\mu={\rm D}u^\mu/\d\tau\equiv u^\mu_{\ ;\nu}u^\nu=
(-R_0/a^2,0,0,0)$. \ Obviously, $\tau$ is the proper time since \
$u_\mu u^\mu=-1$, \ and  the 4-acceleration is constant with modulus
 \begin{equation}
  |A|\equiv|\,\dot u^\mu|={R_0\over a\sqrt{a^2-R_0^2}} \>.
 \label{A}
 \end{equation}
 Since \ $\,u_\mu \dot u^\mu=0$, \ this constant value of $|A|$ is
identical to the modulus of the 3-acceleration measured in the natural
local orthonormal frame of the observer.

The worldlines (\ref{worldlines}) therefore represent the motion of {\it
uniformly} accelerating observers in a de Sitter universe. When \ $R_0=0$ \
the worldline is a geodesic with zero acceleration. On the other hand,
when \ $R_0=a$, \ the acceleration $A$ is unbounded and motion is along the
null cosmological horizon. In general, these uniformly accelerated trajectories
are given by constant values of $Z_2$, $Z_3$  and $Z_4$ on the de~Sitter
hyperboloid, and coincide with the orbits of the isometry generated by the
Killing vector~$\partial_T$.

It is also instructive to visualise these accelerated observers in the
standard global coordinate representation of de~Sitter space:
  \begin{eqnarray}
 &&Z_0=-a\>{\cos\eta\over\sin\eta}\, \>, \nonumber\\
 &&Z_1=a\>{\cos\chi\over\sin\eta} \>, \nonumber\\
 &&Z_4=a\>{\sin\chi\over\sin\eta}\,\cos\Theta \>, \label{global}\\
 &&Z_2=a\>{\sin\chi\over\sin\eta}\,\sin\Theta\cos\Phi \>, \nonumber\\
 &&Z_3=a\>{\sin\chi\over\sin\eta}\,\sin\Theta\sin\Phi \>, \nonumber
 \end{eqnarray}
 where $\eta\in[0,\pi]$, $\chi\in[0,\pi]$. With this, the metric is
 \begin{equation}
 \d s^2 ={a^2\over\sin^2\eta} \left[ \d\chi^2
+\sin^2\chi(\Theta^2+\sin^2\Theta\,\d\Phi^2) -\d\eta^2\right] \>.
\label{conffmetr}
 \end{equation}
 This immediately leads to the familiar conformal diagram of de~Sitter
space (Fig.~1). Comparing expressions for $Z_4$ in (\ref{static}) and
(\ref{global}), it can be seen that the trajectories of accelerated
observers are given by
 \begin{equation}
 \sin\chi={a|A|\over\sqrt{1+a^2A^2}}\>\sin\eta \>,
 \label{globcoordtraj}
 \end{equation}
 and indicated in Fig.~1.

\section{New accelerating coordinates for de Sitter space}

We now introduce a new coordinate system which is particularly well
suited for studying uniformly accelerating point sources in de~Sitter
space. This is given by the following parametrization of (\ref{fourhyp})
  \begin{eqnarray}
 &&Z_0={\sqrt{a^2-r^2}\,\sinh(T/a) \over
     \sqrt{1+a^2A^2} + A\,r\cos\theta} \>, \nonumber\\
 &&Z_1=\pm{\sqrt{a^2-r^2}\,\cosh(T/a) \over
     \sqrt{1+a^2A^2} + A\,r\cos\theta}\>, \nonumber\\
 &&Z_4={\sqrt{1+a^2A^2}\,r\cos\theta+a^2A\over
     \sqrt{1+a^2A^2} + A\,r\cos\theta} \>, \label{acceler}\\
 &&Z_2={r\sin\theta\cos\Phi \over
     \sqrt{1+a^2A^2} + A\,r\cos\theta}\>, \nonumber\\
 &&Z_3={r\sin\theta\sin\Phi \over
     \sqrt{1+a^2A^2} + A\,r\cos\theta}\>. \nonumber
 \end{eqnarray}
 With \ $r\in[0,a]$, \ $T\in(-\infty,\infty)$, \ $\theta\in[0,\pi]$, \
$\Phi\in[0,2\pi)$, \ this again covers two causally disconnected areas of
the de~Sitter manifold. In these coordinates, the de~Sitter space is
represented in the form
 \begin{eqnarray}
 &&\d s^2 = {\ 1\over[\sqrt{1+a^2A^2} + A\,r\cos\theta]^2}
 \Bigg\{{\d r^2\over1-{r^2\over a^2}} \nonumber \\
 &&\qquad\quad +r^2(\d \theta^2+\sin^2\theta\,\d\Phi^2) -
 {\textstyle \left(1-{r^2\over a^2}\right)\d T^2\ }\Bigg\} \>.
 \label{accmetr} 
 \end{eqnarray}
This new static metric is conformal to the standard form
(\ref{staticmetr}) to which it reduces when $A=0$. The transformation
between (\ref{staticmetr}) and (\ref{accmetr}) is given by
relating $r,\theta$ to $R,\Theta$ as
 \begin{eqnarray}
 &&R^2-a^2={{r^2-a^2} \over
     [\sqrt{1+a^2A^2} + A\,r\cos\theta\,]^2} \>, \nonumber\\
 &&R\sin\Theta={r\sin\theta \over
     \sqrt{1+a^2A^2} + A\,r\cos\theta}\>, \label{rtheta}\\
 &&R\cos\Theta={\sqrt{1+a^2A^2}\,r\cos\theta+a^2A\over
     \sqrt{1+a^2A^2} + A\,r\cos\theta} \>,  \nonumber
 \end{eqnarray}
 which can be obtained by comparing (\ref{static}) with (\ref{acceler}).

It is obvious that the origin \ $r=0$ \ of the coordinates
(\ref{accmetr}) corresponds to \ $R_0=a^2|A|/\sqrt{1+a^2A^2}$ \ and  \
$Z_2=0=Z_3$. \ Substituting the above value of $R_0$ into (\ref{A}), we
conclude that the parameter $A$ in the metric (\ref{accmetr}) is exactly
the value of the acceleration of the corresponding observer. Therefore,
{\it the origin \ $r=0$ \ of the coordinates in} (\ref{accmetr}) {\it is
accelerating in a de~Sitter universe with uniform acceleration $A$}. This
uniformly accelerated trajectory corresponds to the motion of {\it two
distinct points} on the de~Sitter hyperboloid, one in each of the two
causally disconnected static regions. When \ $A>0$ \ the coordinate
singularity \ $r=0$ \ is located at \ $\Theta_0=0$ \ so that \
$Z_4=R_0>0$, \ whereas when \ $A<0$ \ it is located at \ $\Theta_0=\pi$ \
so that \ $Z_4=-R_0<0$. \ It may further be noted that all observers
having arbitrary constant values of $r$, $\theta$ and $\Phi$ also move
with uniform acceleration (generally different from~$A$).

The character of these coordinates can be further understood by
expressing them in terms of the global coordinates (\ref{global}) by
 \begin{eqnarray}
  1-{r^2\over a^2} &=& {\sin^2\eta-\sin^2\chi
\over[\sqrt{1+a^2A^2}\,\sin\eta-aA\sin\chi\cos\Theta]^2} \>, \nonumber\\
 r\sin\theta &=& {a\sin\chi\sin\Theta
\over\sqrt{1+a^2A^2}\,\sin\eta-aA\sin\chi\cos\Theta} \>, \nonumber\\
 \tanh\left({T\over a}\right) &=& \mp\> {\cos\eta\over\cos\chi} \>.
 \end{eqnarray}
 Notice again that \ $r=0$ \ implies both (\ref{globcoordtraj})
and \ $\Theta=0$ \ (or \ $\Theta=\pi$ \ for \ $A<0$). \
These coordinates cover the regions I or III according to the two signs
of $Z_1$ in (\ref{acceler}). This is illustrated in the conformal diagram
of de~Sitter space (Fig.~2). It may be observed that \ $r=0$ \ is
represented by {\it two} worldlines located in regions I and III. The
cosmological horizon occurs when $r=a$. The coordinate $r$ in
(\ref{accmetr}) can in fact be extended through this horizon into the
nonstatic regions II and IV, where it becomes timelike. However, other
coordinate charts are required there since $T\in(-\infty,\infty)$
covers only regions I and III between \ $r=0$ \ and the cosmological horizon.
These are obtained by replacing $\sqrt{a^2-r^2}$ in (\ref{acceler}) by
$\sqrt{r^2-a^2}$.

Note that the {\it explicit} spherical symmetry of the coordinate system
has been removed to accommodate a description of uniform acceleration.
However, this introduces difficulties in describing the de~Sitter timelike
and null infinity given by \ $\eta=0,\pi$ \ (or $R=\infty$) \
using the coordinates of (\ref{accmetr}). It may be observed from
(\ref{rtheta}) that points in regions II and IV in Fig.~2 with \
$R>\sqrt{1+a^2A^2}/A$ \ can only be reached for a limited range of the
coordinate $\theta$. Moreover, infinity can be reached at a {\it finite}
value of the timelike coordinate $r$ such that \
$r\cos\theta=-\sqrt{1+a^2A^2}/A$ \ at which the conformal factor in
(\ref{accmetr}) is unbounded. Only in the ``equatorial plane'' \
$\theta=\pi/2$ \ does \ $r=\infty$ \ correspond to \ $R=\infty$.

\section{The generalised C-metric}

Let us now consider the Einstein space described by the line element 
 \begin{equation}
 \d s^2 = {1\over(p+q)^2} \left({\d p^2\over{\cal P}}
+{\d q^2\over{\cal Q}} + {\cal P}\>\d\sigma^2  -{\cal Q}\>\d\tau^2
\right),
 \label{Cmetric}
 \end{equation}
 where
 \begin{eqnarray}
 &&{\cal P}(p) = A^2 -p^2 + 2m\,p^3 -e^2p^4, \nonumber\\
 &&{\cal Q}(q) = -{\Lambda\over3}-A^2 +q^2 + 2m\,q^3 +e^2q^4.
 \label{PQ}
 \end{eqnarray}
 This is contained in the large family of solutions given by Plebanski and
Demianski \cite{PleDem76} (see equation (6.3), and also
\cite{Carter68}--\cite{KSMH80}) in which the ``rotation'' vanishes.  We
note that the linear terms have been removed in (\ref{PQ}) using the shift 
$p\to p+c_0$ and $q\to q-c_0$, where $c_0$ is a constant. Also, the
coefficients of the quadratic terms have been set to unity by a rescaling of
coordinates. (The possibility of different signs for the quadratic terms has
been considered in \cite{Mann97}.)

When the cosmological constant in (\ref{PQ}) vanishes,
(\ref{Cmetric}) reduces to the well-known C-metric which has been
interpreted \cite{KinWal70} as describing two black holes, each of mass
$m$ and charge $e$, which move in opposite directions relative to a
Minkowski background with acceleration $A$ under the action of some
conical singularities. We will argue below that the metric in the form
(\ref{Cmetric},\ref{PQ}) can be considered as the most natural
generalisation of the C-metric which includes a cosmological
constant~$\Lambda$.

Note that in (\ref{PQ}) the acceleration parameter $A$ is introduced in a
different form to that considered previously (e.g. \cite{ManRos95}). We
use a different scaling, and also adapt the constant terms so that
$\Lambda$ appears explicitly in ${\cal Q}$ only. We will demonstrate below
that this is more convenient for physical interpretation.

In order to maintain the correct space-time signature, it is necessary
that ${\cal P}>0$. This places a restriction on the range of $p$.
However, there is no restriction on the sign of ${\cal Q}$ which may
describe both static and nonstatic regions, with horizons occurring when
${\cal Q}=0$.

Let us assume that there is a finite range of $p$, bounded by roots $p_1$
and $p_2\,({>}p_1)$, in which ${\cal P}\ge0$. This spans the space-time
which we wish to investigate. However, it is convenient to introduce the
parameter $\zeta$ by \ $p=A\,\zeta$, \ so that \
$\zeta_1\le\zeta\le\zeta_2$, \ where \ $\zeta_i=p_i/A$.

Now, let us perform the coordinate transformation
 \begin{eqnarray}
 && r=-\sqrt{1+a^2A^2}\>/\,q \>, \nonumber \\
 && \theta=\int_{\zeta_1}{\d \zeta\over\sqrt{
1-\zeta^2+2mA\zeta^3-e^2A^2\zeta^4}}  \>, \label{transf} \\
 && \Phi=A\,\sigma/c \nonumber \\
 && T=\sqrt{1+a^2A^2}\>\tau \>, \nonumber
 \end{eqnarray}
 where $c$ is a constant which can be specified later. With this, the
metric (\ref{Cmetric}) becomes
 \begin{eqnarray}
 &&\d s^2 = {\ 1\over[\sqrt{1+a^2A^2} - A\,r\,\zeta(\theta)\,]^2} \times
 \label{newmetric}\\
 &&\qquad  \Bigg\{{\d r^2\over F(r)} +r^2\big(\,\d \theta^2
+G^2(\theta)\,c^2\d\Phi^2 \big) -
 F(r)\,\d T^2\ \Bigg\} ,\nonumber
 \end{eqnarray}
 where
 \begin{eqnarray}
&& F(r)=1 -{r^2\over a^2} -\sqrt{1+a^2A^2}\,{2m\over r}
+(1+a^2A^2){e^2\over r^2} \>, \nonumber \\
&& G^2(\theta) =1 -\zeta^2(\theta) +2mA\zeta^3(\theta)
-e^2A^2\zeta^4(\theta) \>,
 \label{FG}
 \end{eqnarray}
 and $\zeta(\theta)$ is the inverse function of $\theta(\zeta)$ given by
the integral in (\ref{transf}). It may be seen that, either when \ $A=0$ \
or when both \ $m=0$ \ and \ $e=0$, \ we have \
$\zeta(\theta)=-\cos\theta$ \ so that \ $G(\theta)=\sin\theta$. \ Otherwise
these can be expressed in terms of Jacobian elliptic functions. For
example, when $G^2(\zeta)$ has four distinct real roots \
$\zeta_1<\zeta_2<\zeta_3<\zeta_4$,
 \begin{equation}
 \zeta(\theta) ={\zeta_1(\zeta_4-\zeta_2)
+\zeta_4(\zeta_2-\zeta_1)\>{\rm sn}^2(n\theta\,|\,k)
\over (\zeta_4-\zeta_2) +(\zeta_2-\zeta_1)\>{\rm sn}^2(n\theta\,|\,k) },
 \end{equation}
 where $\,$sn$\,$ is the Jacobian elliptic function (see
\cite{AbrSte65}), with
 $$ n={\textstyle{1\over2}} \sqrt{(\zeta_3-\zeta_1)(\zeta_4-\zeta_2)},
\qquad  k={(\zeta_2-\zeta_1)(\zeta_4-\zeta_3) \over
(\zeta_3-\zeta_1)(\zeta_4-\zeta_2)}. $$

The possible ranges of the coordinates $p$ and $q$ are illustrated in
Fig.~3. Assuming that ${\cal P}(p)$ and ${\cal Q}(q)$ both have four real
roots, there are six possible static space-time regions for all permitted
values. These are bounded by horizons (roots of ${\cal Q}$) or coordinate
singularities (roots of ${\cal P}$). Attention is concentrated here on the
space-time for which \ $p_1<p<p_2$ \ and \ $p+q<0$. \ This covers two
static and two nonstatic regions separated by a cosmological horizon and
by inner and outer black hole horizons. However, these regions are not
entirely covered by the coordinates $r$ and $\theta$, as is also
illustrated.

It may be observed that, although the $r,\theta$ coordinates
appropriately describe the black hole regions and even the cosmological
horizon, they are not well suited to investigating timelike or null
infinity. The restriction $r>0$ corresponds to $q<0$ (\ref{transf}), and
it is immediately obvious from Fig.~3 that the space-time boundary \
$p+q=0$ \ (where the conformal factor in (\ref{newmetric}) is unbounded)
can only be reached for part of the range of $\theta$, and then for finite
values of~$r$.

It may also be observed that the region \ $p_3<p<p_4$, \ $p+q>0$ \ in
Fig.~3 represents the same space-time with the replacements \ $p\to-p$, \
$q\to-q$ \ and \ $m\to-m$.

\section{Physical interpretation}

It is obvious that, when $A=0$ and $c=1$, (\ref{newmetric}) immediately
reduces to the familiar form of the Reissner--Nordstr{\o}m--de~Sitter
black hole solution in which the parameters $m$ and $e$ have the usual
interpretation and the curvature singularity is located at $r=0$.
Moreover, when $A\ne0$, $c=1$ and $m=0=e$, the metric (\ref{newmetric}) is
identical to (\ref{accmetr}) in which $r=0$ corresponds to two uniformly
accelerating points relative to the de~Sitter background. When $m$ and $e$
are small, (\ref{newmetric}) can naturally be regarded as a perturbation of
(\ref{accmetr}). In this way, the metric (\ref{newmetric}) can be
interpreted as describing {\it a pair of charged black holes uniformly
accelerating in a de~Sitter universe}. However, for a finite acceleration
$A$, it follows from (\ref{FG}) that the effective mass and charge of
each black hole are given respectively by $\sqrt{1+a^2A^2}\,m$ and
$\sqrt{1+a^2A^2}\,e$.

The space-times considered here possess the ``boost'' and ``rotation''
symmetries associated with the Killing vectors $\partial_T$ and
$\partial_\Phi$. In the 5-dimensional representation (\ref{fourhyp}) of
the de~Sitter background, these are given by \
$Z_0\partial_{Z_1}+Z_1\partial_{Z_0}$ \ and \
$Z_2\partial_{Z_3}-Z_3\partial_{Z_2}$ \ which are the analogues of those
described in  \cite{BicSch89} for $\Lambda=0$.

Killing horizons (where the norm of the Killing vector $\partial_T$
vanishes) between static and radiative regions of these space-times occur
when $F=0$. Generally, $F(r)$ can have up to four real roots, one of which
must be negative. However, taking $r\ge0$, $F$ can have at most three positive
real roots. In this case, the space-time will include the familiar inner and
outer black hole and cosmological horizons, although it may be noted that their
geometrical properties are altered by the presence of acceleration. Cases
describing accelerating extreme black holes and naked singularities (in which
the roots are repeated or complex) are also included in (\ref{newmetric}) and
(\ref{FG}) for specific ranges of the parameters.

When the acceleration vanishes, the conformal diagrams of all the possible
(spherically symmetric) cases are known (see \cite{BriHay94}). It is
interesting that these diagrams also describe the global structure of
(\ref{newmetric}) in the plane $\zeta=0$ ``orthogonal'' to the direction
of the acceleration, even in the case when $A\ne0$. On this plane
$r=\infty$ corresponds to de~Sitter-like infinity.

For nonzero acceleration, the complete global structure is very
complicated. For small $m$ and $e$, these space-times can be considered as
perturbations of a de~Sitter universe as illustrated in Figs 1 and 2. In this
context, these figures should be regarded as useful schematic pictures rather
than conformal diagrams since \ $r=0$ \ is now a curvature singularity and other
horizons occur. In particular, it may be observed that the space-time
describes the motion of two black holes accelerating in the de~Sitter
background.

As pointed out in section~3, for some range of $\theta$, the de~Sitter
infinity is reached at a finite value of the coordinate $r$ (which is
timelike in this region) given by \ $r\zeta(\theta)=\sqrt{1+a^2A^2}/A$. \
For this value, the conformal factor in (\ref{newmetric}) is unbounded
(corresponding to \ $p+q=0$ \ in (\ref{Cmetric})). For the coordinates of
(\ref{newmetric}), this apparent angular dependence at infinity is
clearly illustrated in Fig.~3 as discussed in section~4. However, it can be
shown that the {\it proper} time required to reach this boundary is always
unbounded.

Let us finally investigate the nature of the singularities at \
$p=p_1=A\,\zeta_1$ \ and \ $p=p_2=A\,\zeta_2$, \ where \
${\cal P}(p_i)=0$, \ i.e. $G=0$. It follows from (\ref{transf}) that the
singularity at \ $\zeta=\zeta_1$ \ corresponds to \ $\theta=0$, \ and \
$\zeta=\zeta_2$ \ to \ $\theta=2K$, \ where $K$ is the ``quarter period''
(the complete elliptic integral of the first kind related to
(\ref{transf})) which is $\pi/2$ when either \ $A=0$ \ or \ $m=0=e$.

Consider 2-dimensional spacelike surfaces on which $r$ and $T$ are
constants and $\theta\in[0,2K]$, $\Phi\in[0,2\pi)$. By comparing the
circumference of a small circle (fixed $\theta$) around the pole
$\theta=0$ with its ``radius'' (segment with fixed $\Phi$), we
find that in general there is a deficit angle
 \begin{eqnarray}
 \delta_1 &=& 2\pi \left[\, 1-\lim_{\theta\to0}\,{c\,G(\theta)\over\theta}
\,\right] = 2\pi \left[\, 1-c\,G'(0) \,\right]  \nonumber \\
 \noalign{\medskip}
&=& 2\pi \left[\, 1 +c(\zeta_1-3mA\zeta_1^2+2e^2A^2\zeta_1^3 )\,\right].
 \end{eqnarray}
 This is finite and also independent of $r$ and $T$. Thus, the
singularity \ $p=p_1$ \ represents a {\it cosmic string of constant
tension} along the ``semi-axis''  \ $\theta=0$. \ However, for any value of
$m$, $e$, $A$ and $\Lambda$, this axis can always be made regular by
putting \ $c=1/G'(0)$. \ In particular, if \ $A=0$ \ or if $m$ and $e$ are
both zero, \ $G=\sin\theta$ \ and the axis is regular when \ $c=1$ \ as
required for  the spherically symmetric Reissner--Nordstr{\o}m--de~Sitter
space-time or for the de~Sitter universe in accelerating coordinates.

For the alternative pole \ $\theta=2K$, \ the deficit angle of a cosmic
string in the opposite direction is given by
 \begin{eqnarray}
 \delta_2 &=& 2\pi \left[\,
1-\lim_{\theta\to2K}\,{c\,G(\theta)\over2K-\theta} \,\right]
= 2\pi \left[\, 1+c\,G'(2K) \,\right] \nonumber \\
 \noalign{\medskip}
&=& 2\pi \left[\, 1 -c(\zeta_2 -3mA\zeta_2^2 +2e^2A^2\zeta_2^3 )\,\right].
 \end{eqnarray}
 This could be removed by setting \ $c=-1/G'(2K)$. \ However,
it is not possible in general to remove the strings in both directions
simultaneously unless \ $G'(0)=-G'(2K)$. \ This condition is identically
satisfied for \ $m=0=e$ \ or \ $A=0$ \ with \ $c=1$. \ However, for
small $mA$, a linear perturbation about \ $\zeta_1=-1$ \ and \
$\zeta_2=1$ \ indicates that the two strings cannot be removed
simultaneously. It follows that a physically reasonable ($e<m$)
accelerating black hole in a de~Sitter background must be connected to at
least one conical singularity which may be considered to ``cause'' the
acceleration. This has been observed previously \cite{ManRos95} using a
less appropriate coordinate system.

We have already argued that, relative to a de~Sitter background, the
space-times contain two accelerating black holes. These are connected by
at least one string, localised at \ $\theta=0$ \ and/or \ $\theta=2K$. \
Since each point in Fig.~1 represents a 2-dimensional compact
subspace spanned by $\theta$ and $\Phi$, the strings may be located on
either of the two ``antipodal'' points on all of these ``2-spheres'' and
thus connects one black hole to the other.

In the interpretation \cite{Bon83} of the C-metric with
\hbox{$\Lambda=0$}, it was convenient to express the metric in the Weyl
form. Here it is also possible to transform (\ref{newmetric}) to the
analogous Lewis--Papapetrou form, at least within each static region, by
putting \ $\rho=\rho_0+\int\d r/ r\sqrt{F(r)}$, \ where 
$\rho_0$ is a suitable constant. However, since this applies only in
disconnected regions of the space-time, it appears to be less convenient
for a global interpretation of these solutions. In fact, the structure of
the two cases with zero or nonzero $\Lambda$ are very different. It is
therefore not surprising that there is no simple reduction of
(\ref{newmetric}) to the C-metric as \ $\Lambda\to0$. \ Nevertheless, it
is again possible to perform a coordinate transformation
 \begin{eqnarray}
  \rho&=&{\textstyle{1\over2}}b\,(\xi^2+t^2-\psi^2)\>, \nonumber \\
  \theta-K&=&b\,\xi\sqrt{\psi^2-t^2} \>, \label{LPtoBS} \\
  T&=&\pm\,a\,\hbox{arctanh}\,(t/\psi) \>, \nonumber
 \end{eqnarray}
$b=$const., with which the metric becomes
 \begin{eqnarray}
 &&\d s^2 ={\cal F}\d\xi^2 + {\cal G}\d\Phi^2  \nonumber\\
 &&\hskip10mm +{\cal F}{(\psi\d\psi-t\d t)^2\over\psi^2-t^2}
 -{\cal H}{(t\d\psi-\psi\d t)^2\over(\psi^2-t^2)^2} \>, \label{BSform}
 \end{eqnarray}
 where
 \begin{eqnarray}
&&{\cal F}={b^2r^2(\rho)(\xi^2+\psi^2-t^2)\over
   [\sqrt{1+a^2A^2} - A\,r(\rho)\,\zeta(\theta)\,]^2} \>, \nonumber \\
&&{\cal G}={c^2r^2(\rho)G^2(\theta)\over
   [\sqrt{1+a^2A^2} - A\,r(\rho)\,\zeta(\theta)\,]^2} \>,
\nonumber\\
&&{\cal H}={a^2F(r(\rho))\over
   [\sqrt{1+a^2A^2} - A\,r(\rho)\,\zeta(\theta)\,]^2} \>. \nonumber
\end{eqnarray}
The form (\ref{BSform}) exhibits the ``boost-rotational'' symmetry
explicitly. It is similar to the analogous form for $\Lambda=0$. It also
enables a natural extension of the static Lewis--Papapetrou metric to
regions with \ $|t|>|\psi|$.

Note finally that some interesting new features occur when \
$\Lambda\ne0$. \ These arise since the universe is now closed and
expanding. For example, a cosmic string starting at one black hole must
extend to the other. Then, proceeding from the opposite pole of the second
black hole, a second (possible) string eventually returns to the first
black hole from the opposite direction. As a second observation, we note
that the acceleration of any object can have both positive and negative
signs simultaneously. This arises since a point moving away in one
direction is also approaching from the opposite side of the universe.

\section*{Acknowledgements}

We are grateful to J. Bi\v{c}\'ak for some useful suggestions.
This work was supported by a visiting fellowship from the Royal Society
and, in part, by the grant GACR-202/99/0261 of the Czech Republic and
GAUK~141/2000 of Charles University.

%\newpage
%{\LARGE Figure Captions}

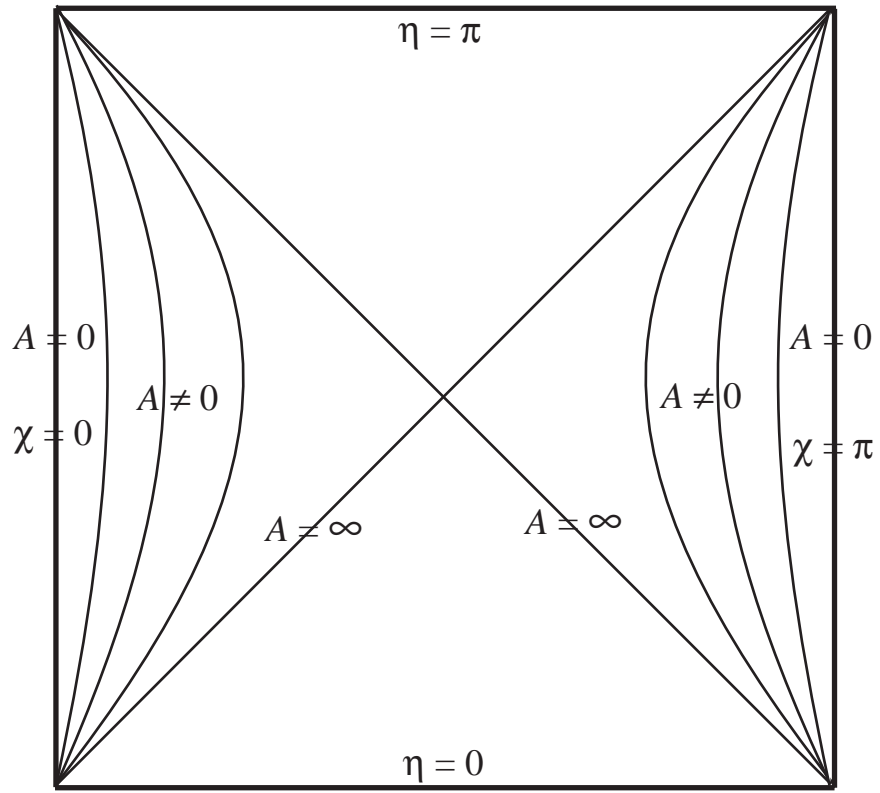
\begin{figure}
\caption{The conformal diagram for de~Sitter space indicating
trajectories of observers with uniform acceleration $A$. Each point
represents a complete 2-sphere. }
\label{Fig.1}
\end{figure}

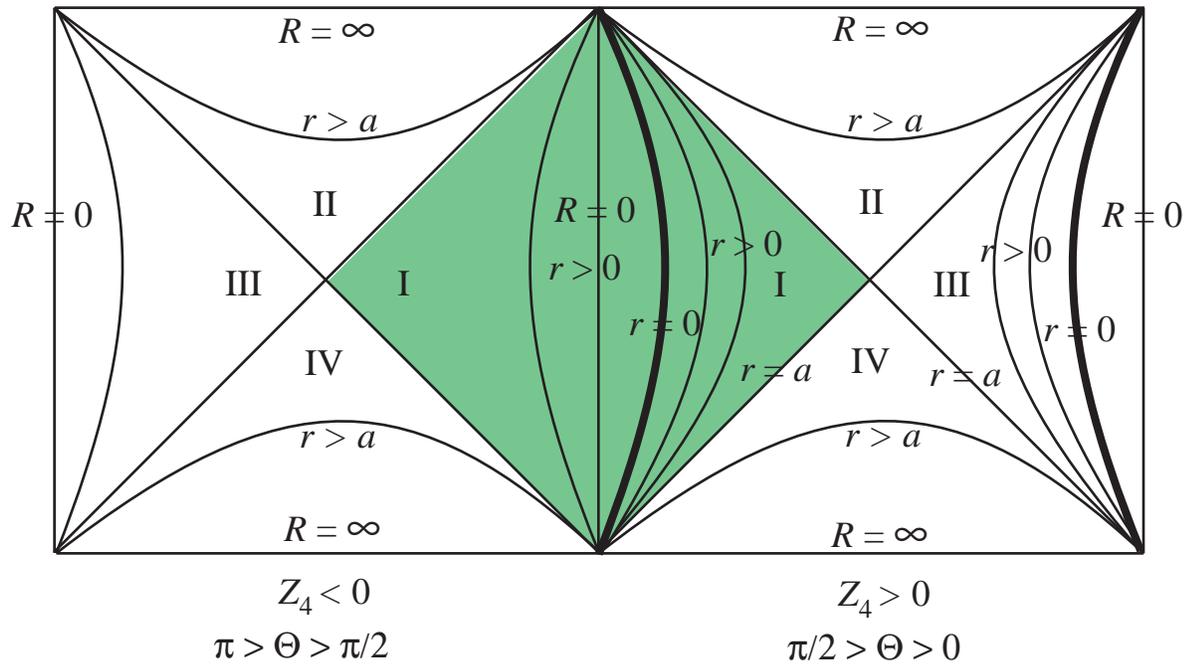
\begin{figure}
\caption{The conformal diagram for de~Sitter space such that each point
represents a hemisphere. The shaded area indicates the region~I covered by
the new coordinates of (\ref{accmetr}) for\ $0\le r\le a$.\ The complete
space repeats this area in III and includes the regions II and IV beyond
the cosmological horizons. Thus \ $r=0$ \ represents two uniformly
accelerating points on causally disconnected opposite sides of the
universe. }
\label{Fig.2}
\end{figure}

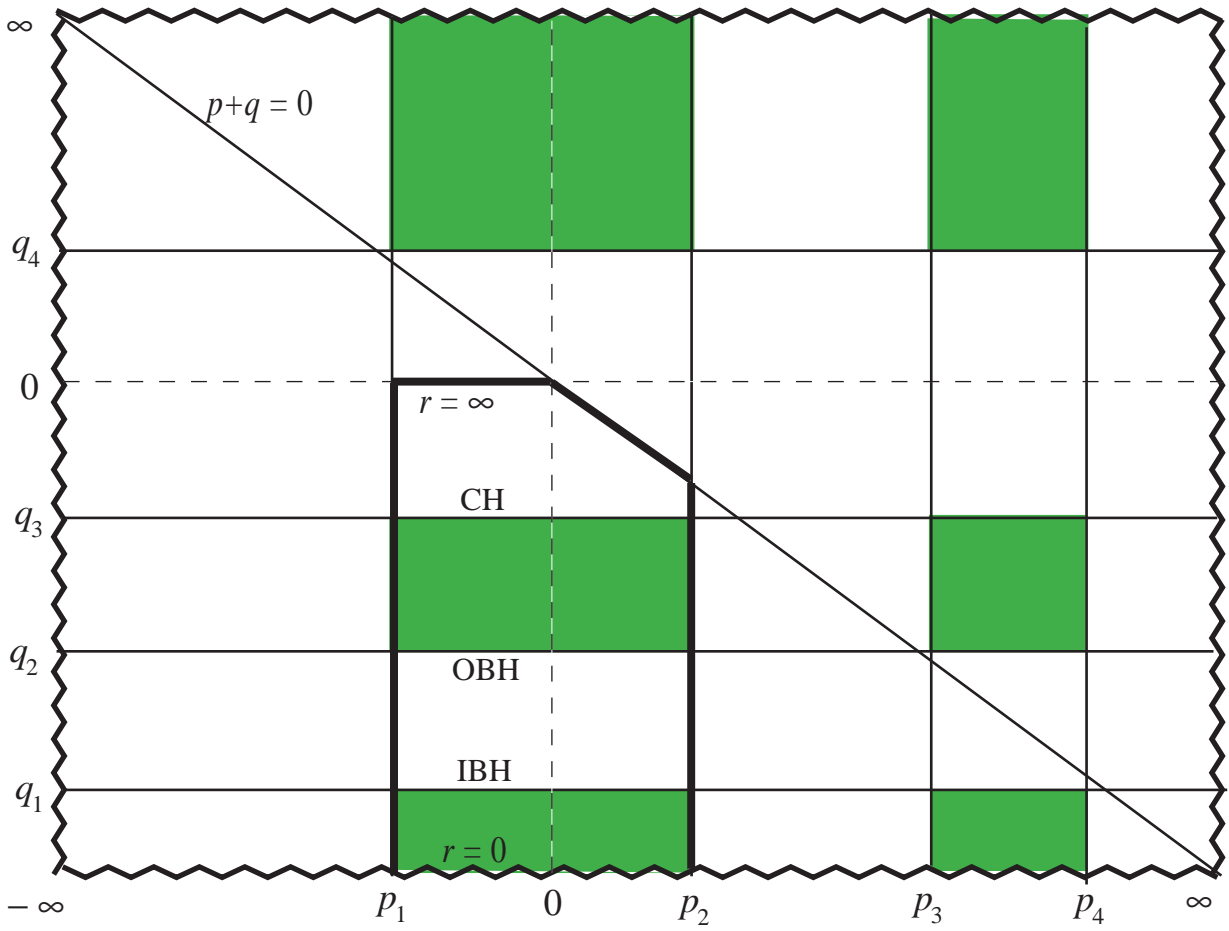
\begin{figure}
\caption{In the full ranges of $p$ and $q$, space-times only occur when
$p$ lies between the roots $p_1$ and $p_2$, or between $p_3$ and $p_4$.
These include six possible static space-time regions which are indicated
by the shaded areas. Attention is focussed here on the space-time spanned
by \ $p_1<p<p_2$ \ and \ $q<-p$. \ This contains two static and two
nonstatic regions separated by a cosmological horizon (CH) and inner and
outer black hole horizons (IBH and OBH). Apart from a region near
conformal infinity, this space-time is also covered by the coordinates $r$
and $\theta$ which span the regions indicated by the area within the bold
lines.}
\label{Fig.3}
\end{figure}

\end{document}